# Optical manipulation of nuclear spin by a two-dimensional electron gas

G. Salis, D. T. Fuchs, J. M. Kikkawa and D. D. Awschalom

*Department of Physics, University of California, Santa Barbara CA 93106*

Y. Ohno and H. Ohno

*Laboratory for Electronic Intelligent Systems, Research Institute of Electrical Communication, Tohoku University, 2-1-1 Katahira, Aoba-ku, Sendai 980-8577, Japan*

**Abstract**

Conduction electrons are used to optically polarize, detect and manipulate nuclear spin in a (110) GaAs quantum well. Using optical Larmor magnetometry, we find that nuclear spin can be polarized along or against the applied magnetic field, depending on field polarity and tilting of the sample with respect to the optical pump beam. Periodic optical excitation of the quantum-confined electron spin reveals a complete spectrum of optically-induced and quadrupolar-split nuclear resonances, as well as evidence for $\Delta m = 2$ transitions.

The electron spin degree of freedom in semiconductors is being investigated for promising new applications like spin electronics [1] and quantum computation [2]. Spin resilience during storage and transport is fundamentally important in this regard, and can be studied by optically introducing electron spin imprints and monitoring their coherent dynamics in space and time. This method has shown that n-doping dramatically reduces extra-electronic spin decoherence, extending spin lifetimes and enabling macroscopic spin transport [3]. Nuclear spins have several orders of magnitude longer lifetimes and are thus favorable candidates for storing quantum bits whose entanglement proceeds via their hyperfine coupling with electron spins [4]. Hence, the dynamics of coherent electron-nuclear interactions are of considerable interest. Dynamic nuclear polarization by electron spin is an incoherent thermodynamic process that has been extensively studied in bulk semiconductors [5-7], quantum wells [8-11], and quantum dots [12]. It was recently proposed that the hyperfine field of periodically excited electron spins could effect nuclear magnetic resonance (NMR) [13], and may provide a foundation for coherent nuclear manipulation using optical techniques. However, optically induced resonances in bulk semiconductors occurred at unexpected magnetic fields so that a direct connection with NMR could not be established [13].

Here, time-resolved measurements of electron spin precession provide unambiguous signatures of all-optical NMR in a modulation-doped GaAs quantum well (QW), and enable spatially selective manipulation of nuclear spin through confinement of the tipping field. Resonances are identified from all three host nuclear isotopes, including quadrupolar splittings and nominally forbidden transitions at half the conventional resonance field. Moreover, we observe low field resonances indicating the survival of nuclear coherence on millisecond time scales.

We studied both single and multiple QWs grown by molecular beam epitaxy with various doping densities [14], and here present data for a single, 7.5 nm wide GaAs QW with $4 \cdot 10^{15}$ m$^{-2}$ Si-doping in the Al$_{0.4}$Ga$_{0.6}$As barriers on both sides. The QW confinement along the (110) crystal direction suppresses D'yakonov-Perel spin relaxation so that spin lifetimes are several nanoseconds from 5 K to room temperature [15], comparable to values found in II-VI semiconductor QWs [16]. These spin lifetimes are an order of magnitude longer than in similar (001) GaAs QWs, so that measurements of the spin precession frequency, $\Omega_L = g\mu_B(B+B_n)/\hbar$, provide a sensitive probe of local magnetic fields, $B_n$ ($g$ is the effective electron g-factor, $B$ the applied magnetic field) [13]. In contrast, such 'Larmor magnetometry' is less sensitive in (001) QWs, where we do not observe the nuclear resonances described below for (110) samples.

A train of 100 fs pulses from a mode-locked Ti:sapphire laser is separated into pump (1.2 mW) and probe (80 µW) beams, which are focused to spatially overlap onto the sample with a diameter of ~70 µm and angular separation of 3 degrees. The circularly polarized pump is tuned to the heavy-hole QW absorption and generates spin-polarized carriers in the QW. We measure the Faraday rotation (FR) of the probe's linear polarization, which is proportional to the electronic magnetization along the probe beam's direction [17]. Modulation of the pump intensity with a mechanical chopper at kHz frequencies allows the use of a lock-in amplifier. Variation of the pump-probe time interval $\Delta t$ reveals the dynamics of electron spins, and reflects both Larmor precession and transverse spin relaxation. Measurements are performed at temperatures $T$ between 5 and 80 K in a magneto-optical cryostat with $B$ applied perpendicular to the pump beam and the $[\bar{1}10]$ crystal direction. The QW can be rotated to adjust the angle $\alpha$ between the [001]-direction and $B$ [inset, Fig. 1(a)].

Nuclei in the QW layer are spin polarized by optical pumping, and their moment is monitored versus magnetic field, temperature, pump-intensity and laboratory time by time-resolved measurement of the electron spin precession. We find that the nuclei are highly polarized and, by tilting the sample, the sign of the polarization with respect to the applied magnetic field can be controlled. Figure 1(a) shows FR measurements at $\alpha = 10°$ and $T = 5$ K. For $B = 0$ T the FR decays over a timescale larger than 2 ns. In an applied field, the FR oscillates as a function of $\Delta t$ and can be fit by an exponentially decaying harmonic oscillation added to a non-oscillatory exponential decay. Both the decay times and the oscillation frequency $\Omega_L$ depend strongly on the sign of $B$ and on the duration of prior sample exposure to the pump beam. Starting with a non-illuminated sample, $\Omega_L$ saturates exponentially with a time constant of several minutes at 5 K, indicating a nuclear origin of the $B$-field asymmetry. To avoid transient effects, samples are saturated for 10 minutes in the optical fields prior to data collection. At $T = 5$ K, we investigate the field asymmetry as a function of $\alpha$ [Fig. 1(b)]. Values for $\Omega_L$ are obtained from fits to FR scans measured at $B = 6$ T and $-6$ T. Although $\Omega_L$ is not symmetric in $\alpha$ and $B$, its value is not changed under inversion of both $\alpha$ and $B$. This same symmetry is observed when the helicity of the pump polarization and one of the two variables $\alpha$ and $B$ is inverted (not shown).

In a greyscale map of the FR data vs. $B$ and $\Delta t$, Fig. 1(c) and (d) show the $B$-dependence of the Larmor precession for $\alpha = 10$ and $0°$ at $T = 5$ K. At $\alpha = 10°$, the field-asymmetry persists over the entire field range. Neither at positive nor negative fields can the data be described by a Zeeman splitting where $\Omega_L$ increases linearly with $B$. At $\alpha = 0$, $\Omega_L$ is symmetric in $B$, as expected from symmetry reasons. However, the FR changes in a complicated way with $B$, and the spin precession cannot be described by a single frequency.

The data suggests that $\Omega_L$ is strongly influenced by a local magnetic field $B_n$ originating from a hyperfine coupling $A\mathbf{I} \cdot \mathbf{S}$ between electron spin $\mathbf{S}$ and dynamically polarized nuclear spin $\mathbf{I}$. The hyperfine constant $A$ contains the squared modulus of the electron wavefunction at the position of the nuclei. An average nuclear spin polarization $\langle \mathbf{I} \rangle$ gives rise to an effective nuclear field $\mathbf{B}_n = A\langle \mathbf{I} \rangle / g\mu_B = b_n \langle \mathbf{I} \rangle / I$. In bulk GaAs, $b_n = -5.3$ T was predicted [6]. Quantum confinement and doping reduces $g$ [18], thus $B_n$ can attain even higher values.

In order to extract the nuclear field $B_n$, we identify the nominal electron Zeeman splitting $g\mu_B B$ by reducing the pump intensity, which diminishes dynamic nuclear polarization. Figure 2(a) shows the intensity dependence of $\Omega_L$ for various temperatures at $\alpha = 10°$. By reducing the pump intensity, $\Omega_L$ asymptotically converges to the same value for $B = 6$ and $-6$ T. This value reflects the Zeeman splitting, which amounts to 28 GHz at 5 K, corresponding to $|g| = 0.053$. The same g-factor is found at $B = 3$ T. The g-factor decreases with temperature and reaches 0.043 at 80 K. Subtracting the Zeeman frequency from $\Omega_L$ reveals the strength of the nuclear field. In Fig. 2(b), we plot $B_n = \hbar \Omega_L / |g|\mu_B - |B|$ for temperatures between 5 K and 80 K. The sign of $B_n$ reflects the relative sign of nuclear polarization with respect to the applied field. $B_n$ is negative for $B < -2$ T and positive for $B > 0$. Spin precession cannot be resolved between $-2$ T and - 0.3 T. $B_n$ disappears at $B = 0$ and increases rapidly with $|B|$ for $|B| < 0.3$ T and $T < 20$ K [not resolved for negative $B$ in Fig. 1(c)]. At 5 K and $B = 2$ T, $B_n$ peaks at 11 T. At $T = 80$ K, $B_n$ still reaches 0.4 T.

Dynamic nuclear polarization occurs when the electron spin is driven from equilibrium and attempts to thermalize through the hyperfine interaction (Overhauser effect). Tilting the sample redirects the pump propagation in the sample according to Snell's Law, generating electron spins with a longitudinal component. This component, when sufficiently large compared to the thermal electron spin polarization, determines the direction of the dynamically polarized nuclear spin. Such reasoning explains the asymmetry of nuclear polarization with $B$, α, and pump helicity, but does not account for the disappearance of nuclear polarization at $B = 0$, its increase at small negative fields or the disappearance of FR oscillations between –2 T and –0.3 T.

To explain this low-field behavior, the effect of the electron hyperfine field, $\mathbf{B}_e \propto \langle \mathbf{S} \rangle$, on the nuclear spin, and of $\mathbf{B}_n$ on the electron spin, has to be taken into account. Here, $\langle \mathbf{S} \rangle$ is the time-averaged electron spin, whose component perpendicular to $\mathbf{B} + \mathbf{B}_n$ is reduced by spin-precession. Because the direction of $\mathbf{B}_n$ can deviate substantially from $\pm\mathbf{B}$ if $\mathbf{B}_e$ is strong enough, a situation where $\mathbf{B} + \mathbf{B}_n$ points along the pump direction can occur, in which case no electron spins precess, and a longitudinal electron-spin decay is measured. This could explain the observed fading of the oscillation amplitude at $-2$ T $< B < -0.3$ T. It is beyond the scope of this paper to establish a theory for the field-dependence of $\langle \mathbf{I} \rangle$. The saturation and subsequent decrease of $B_n$ for large fields can be qualitatively explained by considering nuclear spin-diffusion and additional nuclear spin relaxation mechanisms [19].

The huge $B_n$ achievable in our sample yields high sensitivity to resonant nuclear depolarization induced by the pump laser pulse train with repetition rate of $\nu = 76$ MHz. In this scheme [13], the periodically excited electron spin generates an electron hyperfine field $\mathbf{B}_e$ located in the QW and modulated at frequency $\nu$. Under resonant conditions, $\mathbf{B}_e$ can act as a tipping field for the respective nuclei, leading to a depolarization of the nuclear spin, a reduction in $B_n$, and a change of $\Omega_L$.

For the detection of all-optical NMR we fix the pump-probe delay $\Delta t$ at 450 ps and sweep $B$ from 7.5 to 2 T with a sweep rate of 50 mT/min at $\alpha = 5°$ [Fig. 3(a)]. The FR oscillates with $B$ due to its proportionality to $\cos(g\mu_B \Delta t (B+B_n)/\hbar)$. At nuclear resonance, the decrease in $B_n$ leads to a peak in the FR, whose sign depends on the phase of the FR oscillation at $\Delta t$ and the amount $B_n$ is reduced on resonance.

Besides the $^{69}$Ga-resonance at 7.44 T [13], we observe a peak at 5.85 T belonging to $^{71}$Ga and three peaks at 2.93, 3.72 T and 5.21 T. The latter are at half of the expected resonance fields of $^{71}$Ga, $^{69}$Ga and $^{75}$As, respectively. Since no subharmonic patterns exist in the laser pulse train, we consider the possibility of $\Delta m = 2$ transitions ($m$ being the spin quantum number along the field) within the nuclear spin-3/2 levels. Although $\Delta m = 2$ is nominally forbidden for magnetic transitions, we observe quadrupolar resonance splittings consistent with their occurrence. Figure 3(b) shows measured FR taken with slow sweep rates of 1 mT/min, plotted against $B - B_0$, where $B_0$ is the center of the respective resonance. At 3.72 T we observe a doublet, consistent with two possible $\Delta m = 2$ resonances within the four nuclear spin levels [Fig. 3(c), inset]. At the full field, $B_0 = 7.44$ T, we see a superposition of a triplet and a doublet. We speculate that the doublet reflects the two quadrupolar split $\Delta m = 2$ transition driven by the second-harmonic component of periodic excitation, whereas the triplet comprises three $\Delta m = 1$ resonances. Consistent with this picture, the triplet and both doublets all have the same field separation $\Delta B$. The splittings do not depend on the laser intensity, excluding their explanation by the Knight-shift due to partial nuclear overlap with polarized electrons [9]. We find $\Delta B = 4.3$ mT for the $^{69}$Ga resonance. The data for $^{71}$Ga shows the same behavior with $\Delta B = 2.0$ mT [Fig. 3(c)]. The observed doublet of $^{75}$As around 5.21 T has a splitting of $\Delta B = 12$ mT. The relative strengths of these splittings are in agreement with the quadrupolar moments of the three nuclei [20]. In the light of these results, an unidentified resonance peak reported earlier in bulk GaAs [13] is interpreted as a $\Delta m = 2$ resonance. However, no quadrupolar splitting was resolved there. Because $\Delta m = 2$ is not a selection rule for magnetic-dipole interactions, mechanisms other than the proposed hyperfine tipping process might underlie the observed all-optical NMR. It is known that $\Delta m = 2$ transitions occur from interactions of the nuclear quadrupolar moment with electric fields [21], which might be periodically modulated by carrier excitation in our experiment.

Additional resonances are found at low fields $B < 20$ mT. In order to minimize the oscillatory background of the electron Larmor precession, we measure the FR at $\Delta t = 12.86$ ns. A narrow peak around $B = 0$ is expected due to additive amplification of electron spin packets from successive pump pulses [22]. Surprisingly, we observe a rich structure within this peak, with features depending on the pump chopper frequency [Fig. 4(a)]. In a greyscale plot showing the derivative of the FR with respect to $B$, this structure is made visible [Fig. 4(b)]. The field-position of the features scales linearly with the chopper frequency, as seen by the connecting lines in Fig. 4(b). This indicates a resonant origin of the complex structure, whose typical feature size is around 1 mT at 6 kHz, on the order of the nuclear gyromagnetic ratio. Such a nuclear resonance involves transverse nuclear-spin relaxation-times on the order of milliseconds and might be due to resonant cooling in the rotating frame [23] rather than NMR.

In conclusion, strong nuclear effects in a (110) QW give rise to a rich spectrum of all-optical NMR for the three spin 3/2 nuclei of the GaAs host, including quadrupolar split resonances and $\Delta m = 2$ transitions. By using QW electrons to resonantly depolarize nuclear spin, the nuclear excitation has been focused to the 7.5 nm wide QW layer. The role of the electron hyperfine field $B_e$ in the process of nuclear spin depolarization needs further consideration. We thank S.E. Barrett for helpful discussions and acknowledge support from the ARO DAAG55-98-1-0366, DARPA/ONR N00014-99-1096, and NSF DMR-9701072. The work at Tohoku University was supported by the Ministry of Education, Japan (# 09244103) and by the Japan Society for the Promotion of Science (JSPS-RFTF97P00202).

**Figure Captions**

FIG. 1. (a) Measured Faraday rotation vs. $\Delta t$ taken at $\alpha = 10°$ and T = 5 K for $B = -6$, 0 and 6 T, after 10 minutes exposure to pump and probe beams. Inset: measurement geometry. (b) Larmor frequency $\Omega_L$ extracted from fits to data as in (a), as a function of $\alpha$ at $B = -6$ T ($\blacklozenge$) and 6 T ($\lozenge$). Greyscale plots of FR vs. $\Delta t$ and $B$ for $\alpha = 10°$ (c) and $0°$ (d) show the field-asymmetry at nonzero angles. For both angles, the Larmor frequency does not increase linearly with $B$.

FIG. 2. (a) $\Omega_L$ as a function of $T$ for different average pump intensities (1.2, 0.37 and 0.12 mW) and for $B = 6$ and $-6$ T, showing decrease of field-asymmetry with higher $T$ and lower pump intensity. The low-intensity data is used to extract $|g|$. (b) Dependence of nuclear field $B_n = \hbar \Omega_L/|g|\mu_B - |B|$ on $B$ and $T$. $B_n$ is aligned (antialigned) with $B$ at $B > 0$ ($B < -2$ T).

FIG. 3. All-optical NMR: Measured FR at $\Delta t = 450$ ps for $\alpha = 5°$ (a). $B$ was swept from 7.5 to 2 T with 50 mT/min. Five distinct peaks are identified as full-field and "forbidden" half-field resonances of the different nuclei. Slow scans (1 mT/min) show quadrupolar splitting $\Delta B$ into triplets (doublets) at full (half) field of the $^{69}$Ga (b) and $^{71}$Ga (c) resonances. The resonances are attributed to $\Delta m=1$ and $\Delta m=2$ transitions (inset). The data at 7.44, 3.72, 5.85 and 2.93 T are taken at $\Delta t = 450$, 410, 490, and 450 ps, respectively.

FIG. 4. (a) FR at $\Delta t$ =12.86 ns for low fields shows structures which depend on the pump chopper frequency (3 kHz and 6 kHz). (b) A greyscale plot of the differential FR as a function of $B$ and the chopper frequency shows various resonances with chopper frequency proportional to $B$.

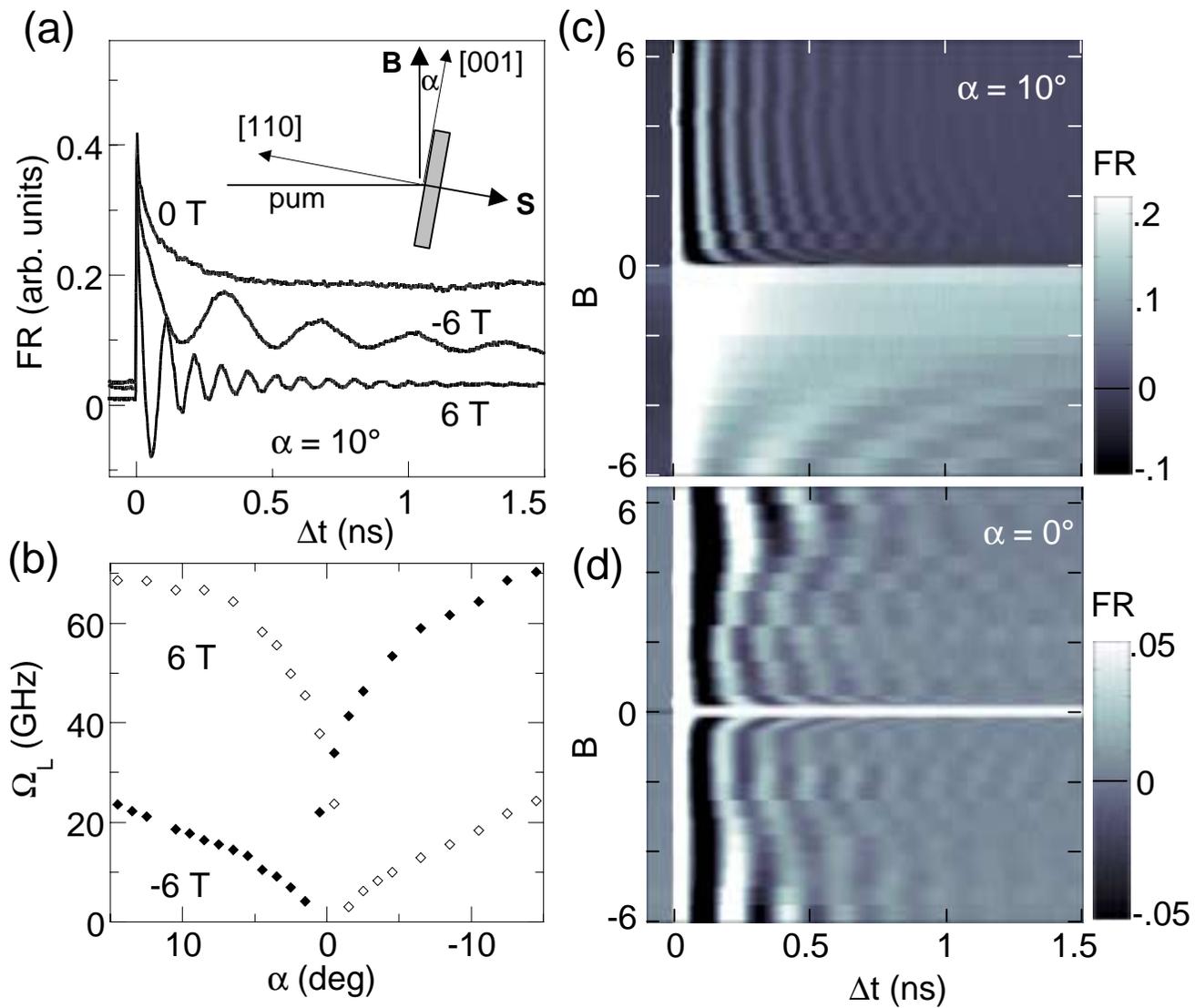

Salis et. al, Fig. 1

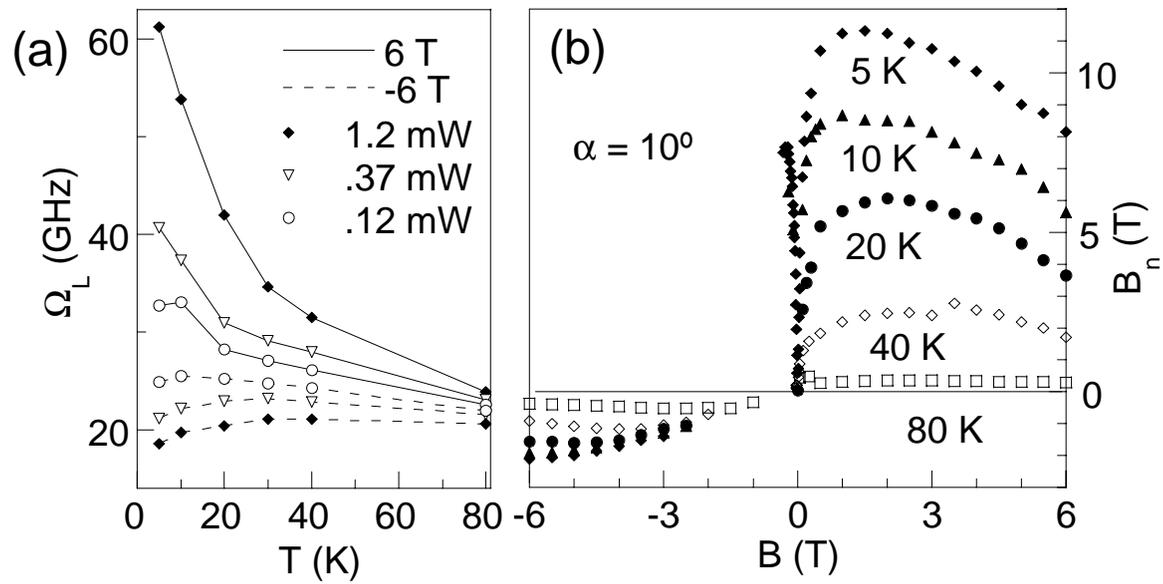

Salis et. al, Fig. 2

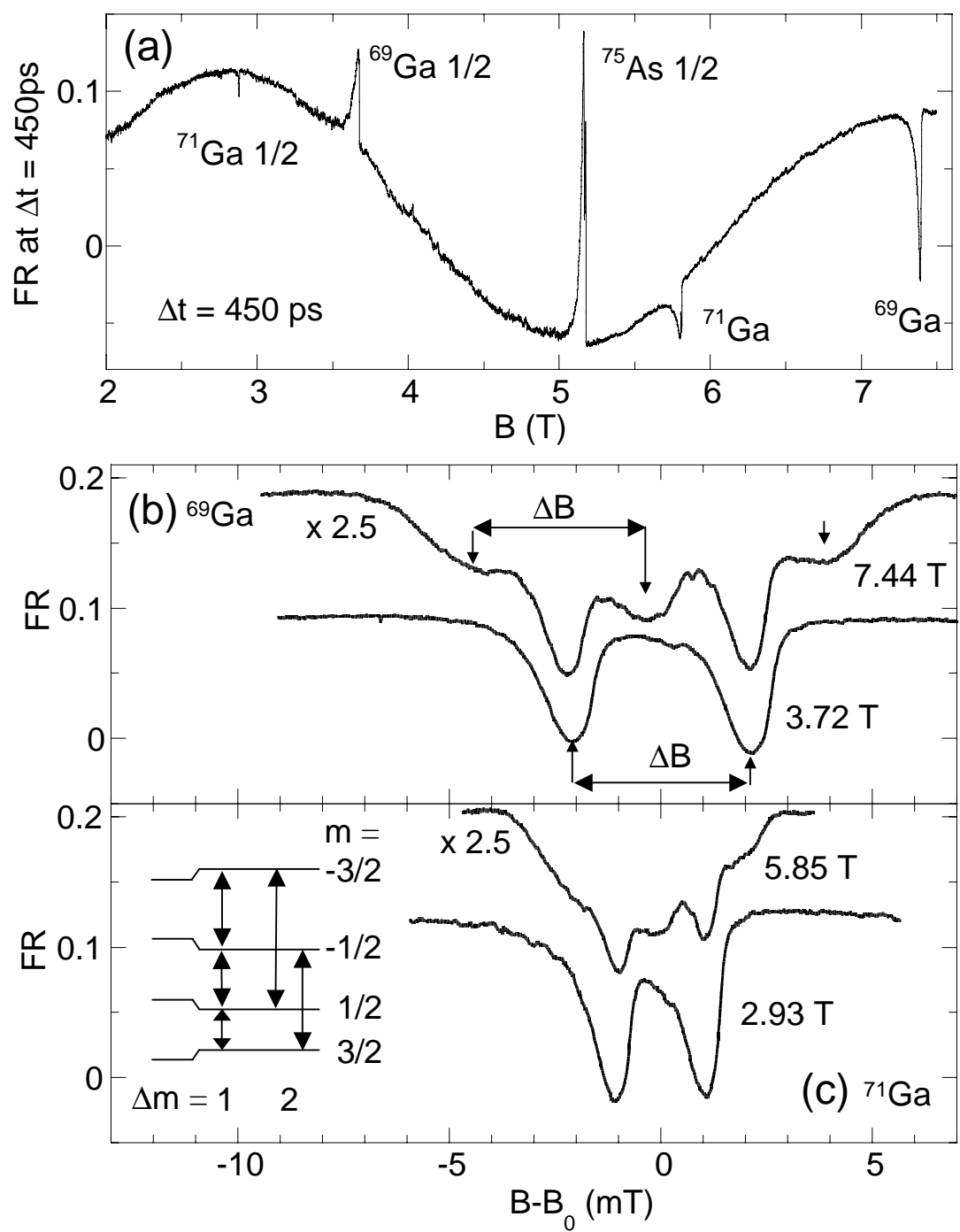

Salis et. al, Fig. 3

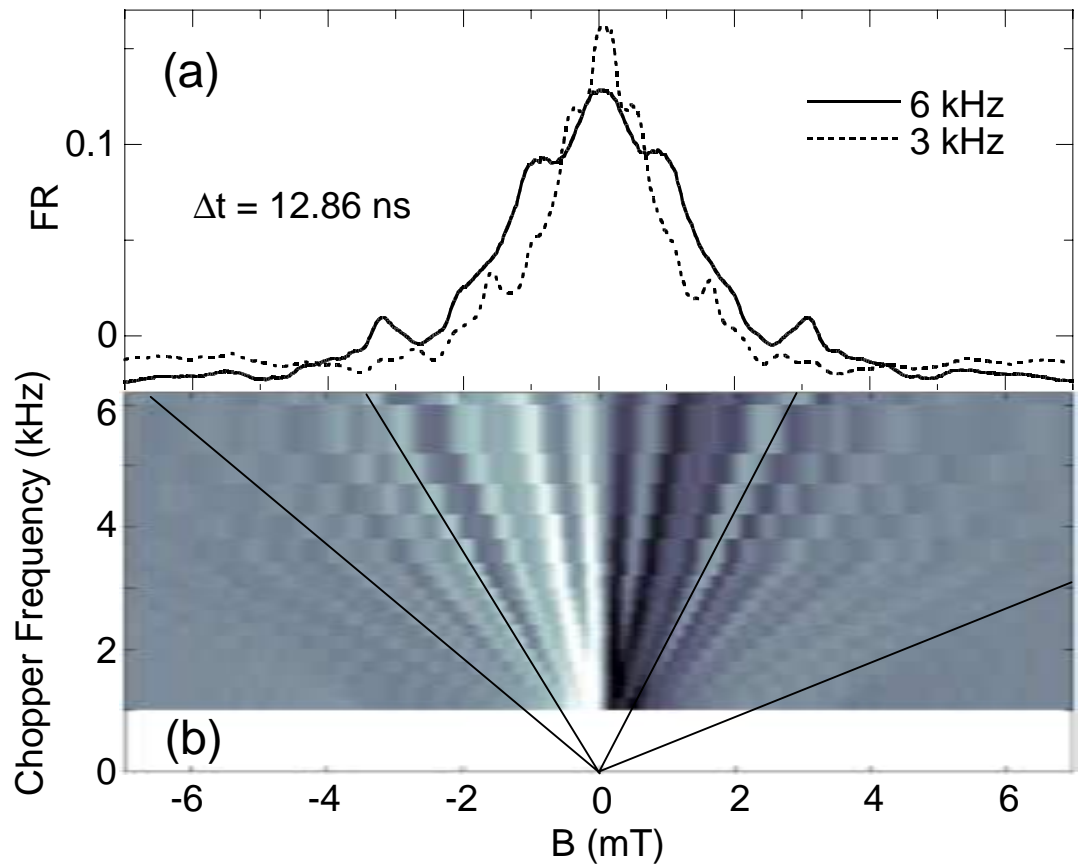

Salis et. al, Fig. 4